\documentclass[useAMS]{mn2e}
\usepackage{lscape}
\usepackage{rotating}
\usepackage{amssymb}
\usepackage{float}

\def\msun{\hbox{M$_\odot$}}

\def\frac{\hbox{f$_{\rm mix}$}}

\def\t4{\hbox{t$_{\rm 4}$}}

\def\cm3{\hbox{cm$^{-3}$}}

\input psfig.sty
\input epsf.sty
\usepackage{graphicx}
\usepackage{epsfig}
\title[A General Problem for Self-Enrichment Scenarios for GCs]
{A General Abundance Problem for All Self-Enrichment Scenarios for the Origin of Multiple Populations in Globular Clusters}
\author[Bastian et al.] {Nate Bastian$^{1}$, Ivan Cabrera-Ziri$^{1,2}$, \& Maurizio Salaris$^{1}$ \\
$^{1}$ Astrophysics Research Institute, Liverpool John Moores University, 146 Brownlow Hill, Liverpool L3 5RF, UK\\
$^{2}$ European Southern Observatory, Karl-Schwarzschild-Stra{\ss}e 2, D-85748 Garching bei M\"unchen, Germany \\
}
\date{Accepted. Received; in original form}
\pagerange{\pageref{firstpage}--\pageref{lastpage}}
\pubyear{2015}
\begin{document}
\maketitle
\label{firstpage}
\begin{abstract}

A number of stellar sources have been advocated as the origin of the enriched material required to explain the abundance anomalies seen in ancient globular clusters (GCs).  Most studies to date have compared the yields from potential sources (asymptotic giant branch stars (AGBs), fast rotating massive stars (FRMS), high mass interacting binaries (IBs), and very massive stars (VMS)) with observations of specific elements that are observed to vary from star-to-star in GCs, focussing on extreme GCs such as NGC~2808, which display large He variations.  However, a consistency check between the results of fitting extreme cases with the requirements of more typical clusters, has rarely been done.  Such a check is particularly timely given the constraints on He abundances in GCs now available.  Here we show that all of the popular enrichment sources fail to reproduce the observed trends in GCs, focussing primarily on Na, O and He.  In particular, we show that any model that can fit clusters like NGC~2808, will necessarily fail (by construction) to fit more typical clusters like 47~Tuc or NGC~288.  All sources severely over-produce He for most clusters.  Additionally, given the large differences in He spreads between clusters, but similar spreads observed in Na--O, only sources with large degrees of stochasticity in the resulting yields will be able to fit the observations. We conclude that no enrichment source put forward so far (AGBs, FRMS, IBs, VMS - or combinations thereof) is consistent with the observations of GCs.  Finally, the observed trends of increasing [N/Fe] and He spread with increasing cluster mass cannot be resolved within a self-enrichment framework, without further exacerbating the mass budget problem.

\end{abstract}
\begin{keywords} galaxies - star clusters, Galaxy - globular clusters
\end{keywords}

\section{Introduction}
\label{sec:intro}

Globular clusters (GCs) are known to display star-to-star abundance variations in specific elements (e.g., He, Na, O, Al) while appearing remarkably homogeneous in other elements (e.g., Si, Ca, Fe - in most clusters).  Most scenarios attempting to explain the (non-)variations in specific abundances invoke ``self-enrichment", where certain stars (a.k.a.~polluters) within a cluster are able to enrich other stars within the same cluster.  These scenarios often invoke multiple star-forming events as a means of getting the enriched material inside other stars.  Due to the unique chemistry observed in GCs, stars undergoing ``hot hydrogen burning" are preferred as polluters.  Popular choices for the polluters are massive asymptotic giant branch stars (AGBs - $5-9$~\msun), fast rotating massive stars (FRMS - $>20$~\msun), interacting massive binary stars (IBs - $>10-20$~\msun), and very massive stars (VMS -$>10^4$~\msun).

However, the pure ejecta from any of the polluting stars mentioned above cannot explain the abundance trends.  Instead, all self-enrichment scenarios include a contribution of ``primordial material", with abundances that match that of the first (original) stars that formed in the cluster (for scenarios that invoke multiple star-forming events, this would be the abundance patterns of the ``1st generation").  In this way, the polluted ejecta is diluted, allowing for a range of abundances to appear within the cluster.  

A straightforward way to test the basic validity of these scenarios is to quantitatively compare the variations in certain elements to others, as these are explicitly linked once the ejecta yields and primordial abundances are chosen.  This is a particularly timely test, due to the constraints on He spreads within GCs that are now available through high precision photometry of main sequence stars in comparison with stellar models (e.g., Milone~2015).  For example, D'Ercole et al.~(2010 - hereafter D10) found that the observed range in the spreads of Na and O in the massive globular cluster NGC~2808, was consistent with the large He spreads inferred from an analysis of its colour-magnitude diagram (CMD, e.g., Piotto et al.~2007), once a modified set of yields for massive AGB stars were adopted.  However, Doherty et al.~(2014) have presented yields of modelled massive AGB stars and have not confirmed the general trends found in the model yields of Ventura et al.~(2013) or those adopted in D10.  These authors find that Na is not enhanced, and in fact, that it is actually destroyed in most models.  This calls into question whether these stars are likely to be contributors to the anomalous abundances observed in GCs.

Cassisi \& Salaris~(2014) and Salaris \& Cassisi~(2014) carried out a similar analysis of four GCs adopting the yields for interacting binary stars of de Mink et al.~(2009) and the early disc accretion scenario of Bastian et al.~(2013b).  The authors adopted simultaneous constraints on Na, O, He and Li, and found that consistent dilution models could be made for NGC~2808 and NGC~104 (47~Tuc), while no consistent model could be found for NGC~6752.  However, despite these and a handful of other studies, quantitative comparisons between observations and theoretical predictions have been lacking in the literature.

In the present work we develop simple dilution models for four popular choices of polluters: AGBs, FRMSs,  IBs, and VMSs.  We quantitatively compare the observed Na and O abundances with model predictions, taking into account constraints obtained for each cluster on the maximum spread in He that is present.  The goal of this study is to test the basic yields for the suggested polluter stars, without discussing the other aspects of each scenario (e.g., the origin of the primordial material, the mass budget problem). The paper is organised as follows; in \S~\ref{sec:obs} we introduce the observational data taken from the literature used in the present work.  In \S~\ref{sec:models} we develop the dilution models and discuss the yields adopted for each type of polluting sources, while in \S~\ref{sec:results} we present our results.  In \S~\ref{sec:discussion} and \S~\ref{sec:conclusions} we discuss our results and present our conclusions, respectively.



\section{Data from the Literature}
\label{sec:obs}

For our analysis we obtained measurements of Na and O for a number of stars in each clusters from the literature.  The data come from the works of Carretta et al.~(2009) and Lind et al.~(2011).  Additionally, for some clusters we made use of the combined catalogue of Roediger et al.~(2014), and refer the reader to that work for references to the original papers and a discussion of the data in detail.

Additionally, in order to better constrain the dilution models, we adopted maximum He spreads, as determined from analyses of CMDs, in particular spreads in the main sequence colours in ultraviolet and optical filters (e.g., Milone et al.~2014).  The reference for each cluster is given in Table~\ref{tab:clusters}.  While the current techniques cannot place strict limits on the absolute abundance of He in clusters, they are very sensitive to He spreads, which is what we exploit in the current work.  We note that if other effects may also affect the CMD that are not taken into account in isochrone modelling (e.g., star-spots and/or strong magnetic fields), the actual He spreads will be less than that inferred from CMD analyses. Hence, the reported He spreads are likely to be upper limits to actual He spreads.

Throughout this work we have assumed that all of the primordial material (for all clusters) had an original He abundance of $Y=0.25$.

Photometric studies (e.g., Piotto et al.~2015) have shown that many GCs display discrete populations in their main-sequences or red giant branches.  Additionally, there is also evidence of discrete populations in some GCs based on their chemical abundances (e.g., Carretta~2014).  This feature may eventually provide a strong constraint when evaluating potential models for the origin of multiple populations, as most models do not account for discreteness naturally, but instead require new parameters and some level of fine-tuning.  However, in the current work, whether or not the ranges of elemental abundances are made up of discrete or continuous distributions will not affect our interpretation.  Throughout this work we discuss ``spreads" in abundances by which we mean either continuous spreads or the difference between discrete populations.


\section{Dilution Models and Yields}
\label{sec:models}

As discussed in \S~\ref{sec:intro} the simplest model for the expected abundance trends in GCs is made by taking yields from the proposed sources of enrichment (i.e. stars from the 1st generation that are providing the enriched ejecta) and combine these with pristine material (i.e., the material with abundances identical to the 1st generation stars).  For each stellar yield (either for stars of different masses, time steps, or forms of the IMF, see below) we mix that yield (for all elements) with the primordial abundance for a variety of fractions, from entirely primordial material ($\frac=0$) to material made up entirely from the ejecta from the polluting stars (e.g., AGB stars, FRMS, interacting binaries), $\frac=1$.  In particular, we trace the expected abundances of Na, O, and He for different mixing fractions.  By just focussing on the yields, coupled with simple dilution models, we will not need to worry about timescales or the physics behind the retention of the ejected/primordial gas and the secondary star-formation within the clusters.  These complications have been discussed elsewhere (e.g., D10; Conroy 
\& Spergel~2011; Bastian et al.~2013a, 2014; Bastian \& Strader~2014; Cabrera-Ziri et al.~2014; 2015; Longmore 2015).

For the AGB scenario we adopt three sets of yields.  The first is taken from D10, which in turn were based on the calculations of Ventura \& DÕAntona (2009) for stars with [Fe/H]$\sim-1.3$.  D10 extrapolated and changed some yields from the Ventura \& D'Antona~(2009) models, based on ``educated guesses", in order to reproduce the abundance trends observed in the ``extreme" stars in NGC~2808.  These extrapolations were done to include ``super-AGB" stars, i.e. AGB stars with masses of $8-9$\msun.

In order to compare with more direct AGB predictions, we also use the yields calculated in Ventura et al.~(2013 - hereafter V13).  As for D10 we adopt the models of [Fe/H]$\sim-1.3$, but we explore the role of metallicity in these calculations in \S~\ref{sec:metallicity}. For stars with masses less than $6$~\msun\ the yields of V13 and D10 are very similar.  However, large differences exist for stars $\geq6.5$~\msun.  The yields of these stars have been adjusted in D10 to match observations of NGC~2808, whereas the actual predictions of the models are used in V13.  The two sets of models can be compared in the top left and right panels of Fig.~\ref{fig:n104}.

We have also investigated the yields of super-AGB stars in the Doherty et al.~(2014) calculations as compared to the D10 and V13 results. Again we adopt the models with [Fe/H]$\sim-1.3$.  These stars produce high He yields ($Y=0.36$) but do not produce significant amounts of Na, remaining near the primordial abundance level.  Doherty et al.~(2014) concluded that these stars are unable to reproduce the extreme stars and also the main populations within the observed clusters.  Due to the clear discrepancy, we do not show the models superimposed with observed GC stars.


Next, we adopt the yields from Decressin et al.~(2007) for the FRMS scenario, who ran models for [Fe/H]$\sim-1.5$.  These authors provide the stellar IMF weighted yields for two different stellar IMFs (with mass function indices of x=1.35 (i.e.,~Salpeter) and x=0.4).  As can be seen from the lower left panel of Fig.~\ref{fig:n104}, the two sets of yields do not differ significantly.

Additionally, we use the expected yields for interacting binaries from the calculations of de Mink et al.~(2009 - hereafter dM09), who adopted [Fe/H]$\sim-1.5$.  We show two sets of yields, one for the ``average yield" from the ejecta from the modelled system, and one for the ``extreme yield" (off the plot), which represents the most extreme yields found in their model.  One strong caveat to these yields is that they are based on a single calculation of one binary system.  The yields from interacting binaries are expected to depend on at least three parameters; 1) the mass ratio of the system, 2) the mass of the primary star, 3) the evolutionary phase when the stars begin interacting (dM09).  Whether or not the yields reported in dM09 are representative of the full range of yields of a full population (sampling the three dimensional parameter space) is currently unclear, and future models will need to address this.  As we will see, the potential for large stochastic variations in the yields is one of the most promising aspects for this particular polluter.

Finally, we test very massive stars (VMS) as potential polluters (Denissenkov \& Hartwick~2014), and use the yields of Denissenkov et al.~(2015).  The enriched material from VMS stars is  extremely deficient in O, while being enhanced in Na and He.  The results are largely similar to that found for the IB and FRMS scenarios and will be discussed in \S~\ref{sec:vms}.

We note that the yields of all the potential sources of enrichment are uncertain at some level (e.g., as seen in the large differences between model AGB yields in V13 and Doherty et al.~2014).  In order to test whether the lack of agreement between the predictions and observations is due to uncertainties in the model yields, or rather reflects a more fundamental problem with self-enrichment scenarios, we will also compare the observations with ``empirical" dilution models, based on the primordial and ``extreme" stars observed in specific clusters.  These tests and results are discussed in \S~\ref{sec:empirical}.

All yields have been scaled from the initial abundance used in the models to the adopted ``primordial abundances" observed in each cluster, i.e., we assumed that the amount of enrichment or depletion was independent of the initial abundances. This was done by shifting the initial abundance of the models (in [Na/Fe] and [O/Fe]) to the ``primordial" value for each cluster.   This results in the same range of abundance variation for each cluster as found in the models.  Such shifts have also been carried out in D10, and result in the most optimistic match between the observations and the models.

Finally, we note that there is still a major open question for many dilution models, namely where does the gas with primordial abundance patterns come from?  This question is beyond the scope of the current paper, where we simply assume that such a reservoir exists, but we note that this may be a fundamental stumbling block for scenarios that invoke multiple episodes of star formation and has been discussed at length in a number of recent works (e.g., Bastian et al. 2014; Bastian \& Strader~2014; Cabrera-Ziri et al.~2015; Hollyhead et al.~2015).

\section{Results}
\label{sec:results}

\begin{table*}
\caption{Observations used in the current work.  [Fe/H] values were taken from Harris~(1996; the 2010 version).  Note that Roediger et al.~(2014) is a compilation of data from a number of papers.  We refer the interested reader to that paper for more details.  $^1$These studies place independent constraints on the He spread based on the Horizontal Branch morphology of the clusters. $^2$We note that Larsen et al.~(2015) place a significantly lower limit on the maximum He spread within NGC~7078 (i.e., $\Delta(Y)_{\rm max} < 0.03$).}
\label{tab:clusters}
\begin{tabular}{lcccccc}
\noalign{\smallskip}
\hline
\hline
\noalign{\smallskip}
Cluster &  $\Delta(Y)_{\rm max}$ & [Fe/H]  & [Na/Fe] and [O/Fe] & $\Delta(Y)$\\
 & & & reference & reference \\
\hline
\noalign{\smallskip}
NGC~104 & 0.03 & -0.72 & Roediger et al.~(2014) & di Criscienzo et al.~(2010), Milone et al.~(2012c); Gratton et 
al.~(2013)$^1$\\
NGC~288 & 0.013 & -1.32 & Carretta et al.~(2009)& Piotto et al.~(2013)\\
NGC~2808 & 0.14 & -1.14& Roediger et al.~(2014) & Milone et al.~(2012b); Dalessandro et al.~(2011)$^1$\\
NGC~6121 & 0.04 & -1.16 & Carretta et al.~(2009) &  Villanova et al.~(2012)\\
NGC~6397 & 0.01 & -2.02& Lind et al.~(2011) & di Criscienzo et al.~(2010), Milone et al.~(2012a)\\
NGC~6752 & 0.035 & -1.54 & Roediger et al.~(2014) & Milone et al.~(2013) \\
NGC~7078 & 0.053$^2$ & -2.37 & Roediger et al.~(2014) & Milone et al.~(2014); Milone et al.~(in prep.)\\
NGC~7099 & 0.030 & -2.26 & Carretta et al.~(2009) & Mucciarelli et al.~(2014)\\
\hline
\end{tabular}
\end{table*}

\subsection{AGB scenario}

The basic idea of the AGB scenario is that the ejecta of AGB stars of a first generation within a massive cluster cannot escape the gravitational potential well.  This ejecta cools and fall to the centre of the cluster.  The cluster then accretes material that matches the abundances of the first generation which mixes with the AGB ejecta and forms subsequent (second and further) generations.  The timing of the re-accretion of pristine material is critical (e.g., D10, Cabrera-Ziri et al.~2015), and thus far, it is not clear where this material comes from\footnote{Conroy \& Spergel~(2011) suggest that clusters retain $\sim10$\% of their initial stellar mass in gas, which acts as a net, sweeping up material from the ISM as the cluster moves through the galaxy.  However, this process would not match the observed abundance spreads in GCs (cf. D'Ercole et al. 2011). Furthermore, clusters which are above the mass limit where this effect is expected, do not have such gas/dust reservoirs (Bastian \& Strader~2014; Cabrera-Ziri et al.~2015).}.

Here, we use the yields of three sets of calculations for massive and super-massive AGB stars as input for our dilution model (see \S~\ref{sec:models}).  

\subsubsection{Yields from D'Ercole et al.~2010}
\begin{figure*}
\centering
\includegraphics[width=8cm]{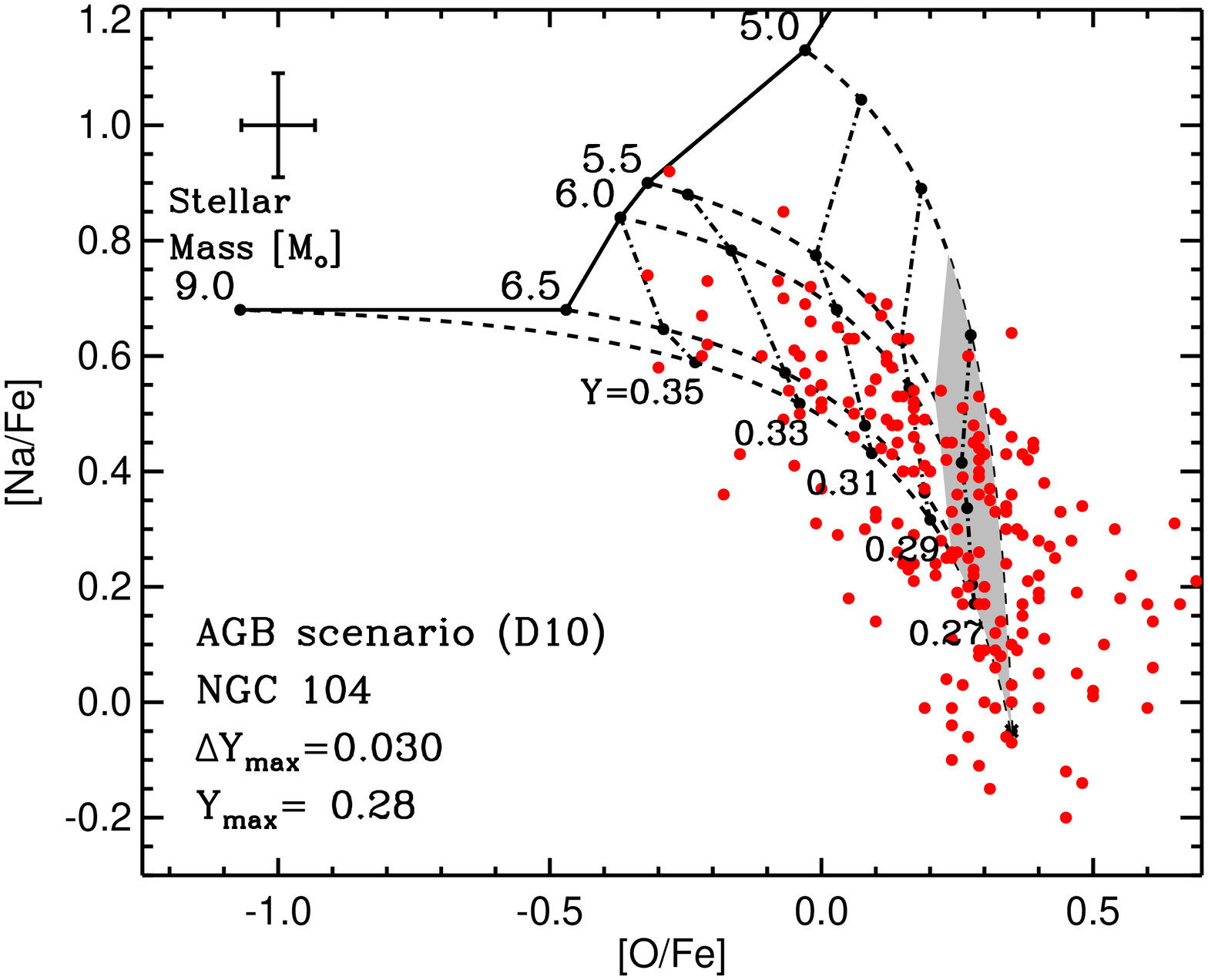}
\includegraphics[width=8cm]{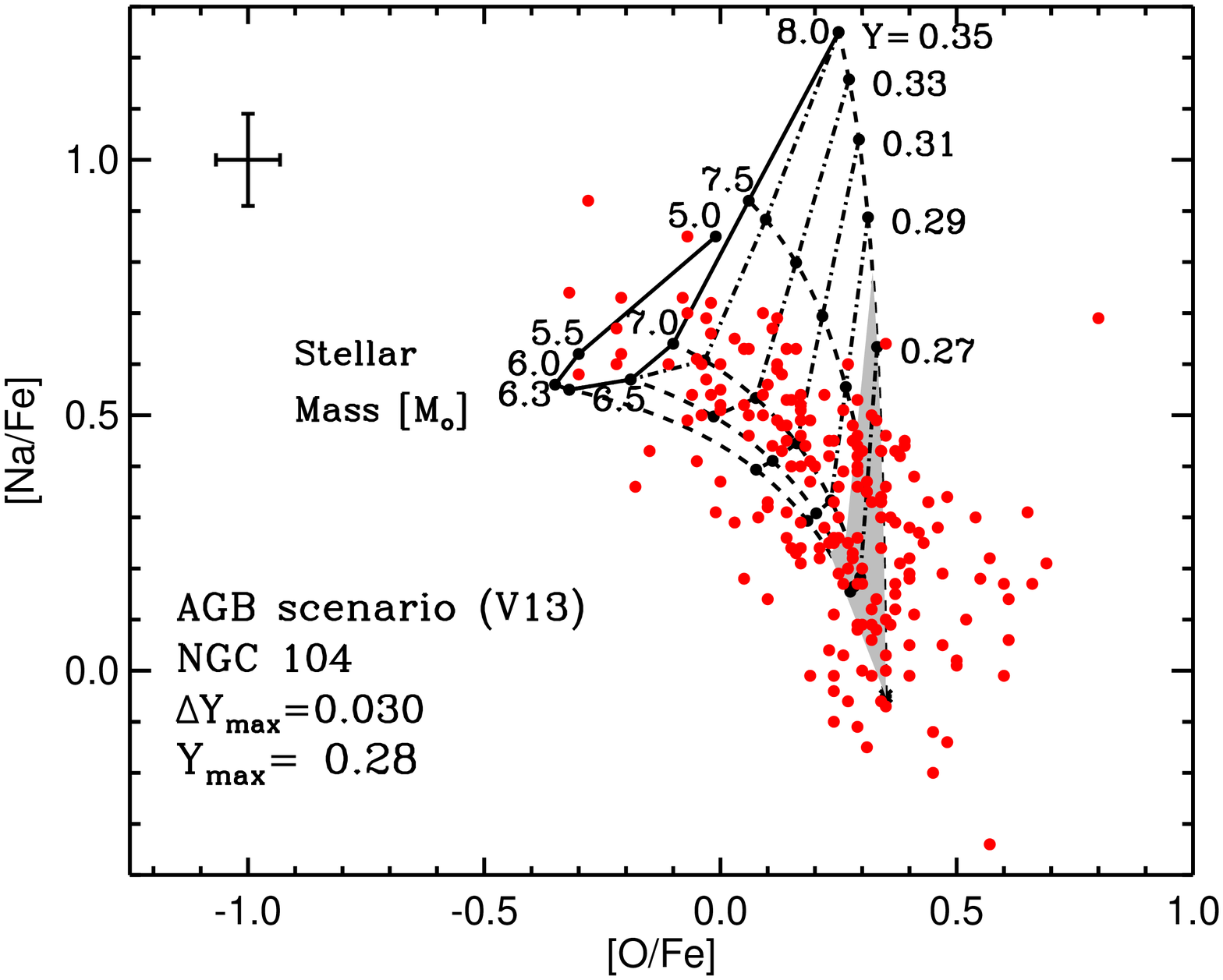}
\includegraphics[width=8cm]{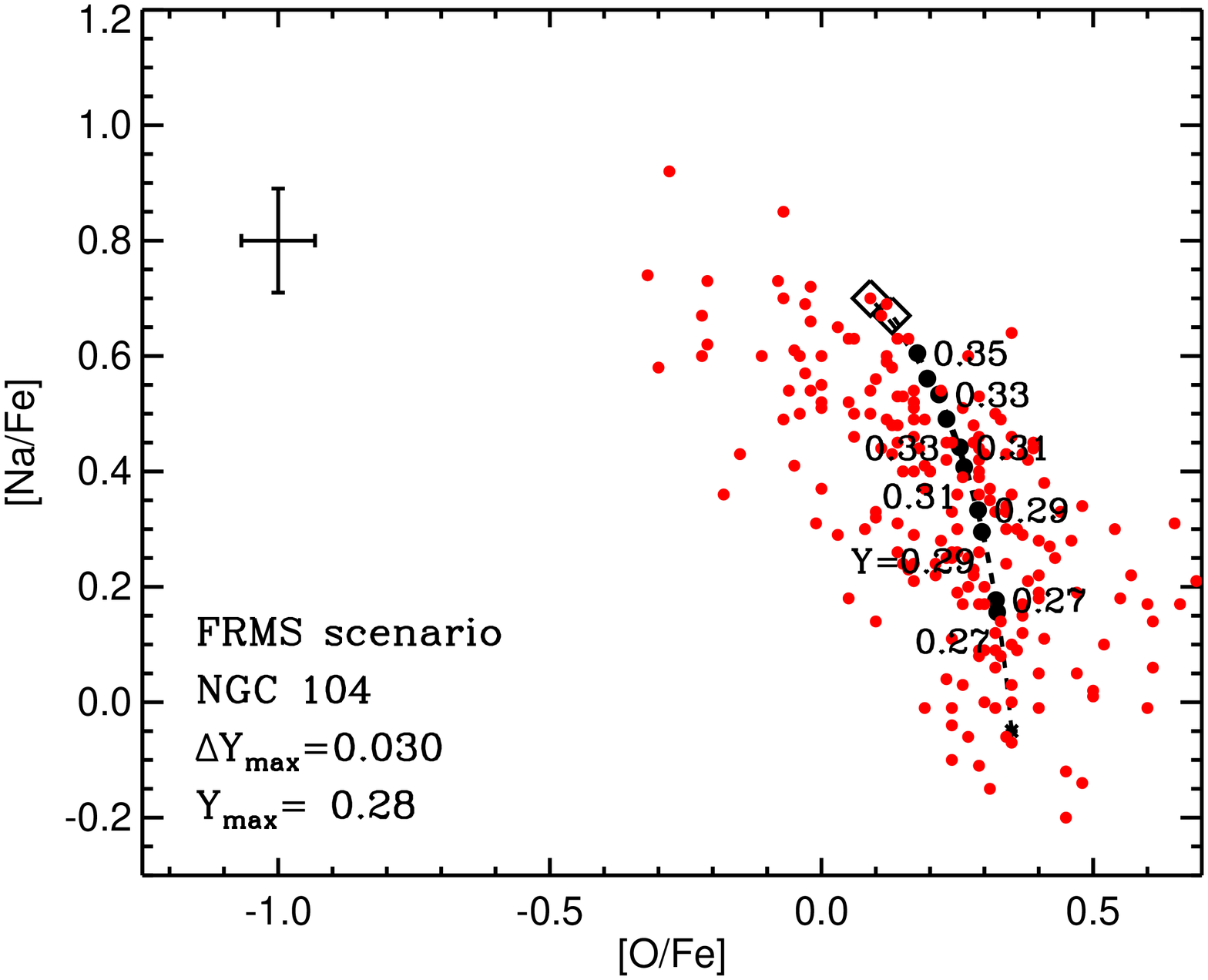}
\includegraphics[width=8cm]{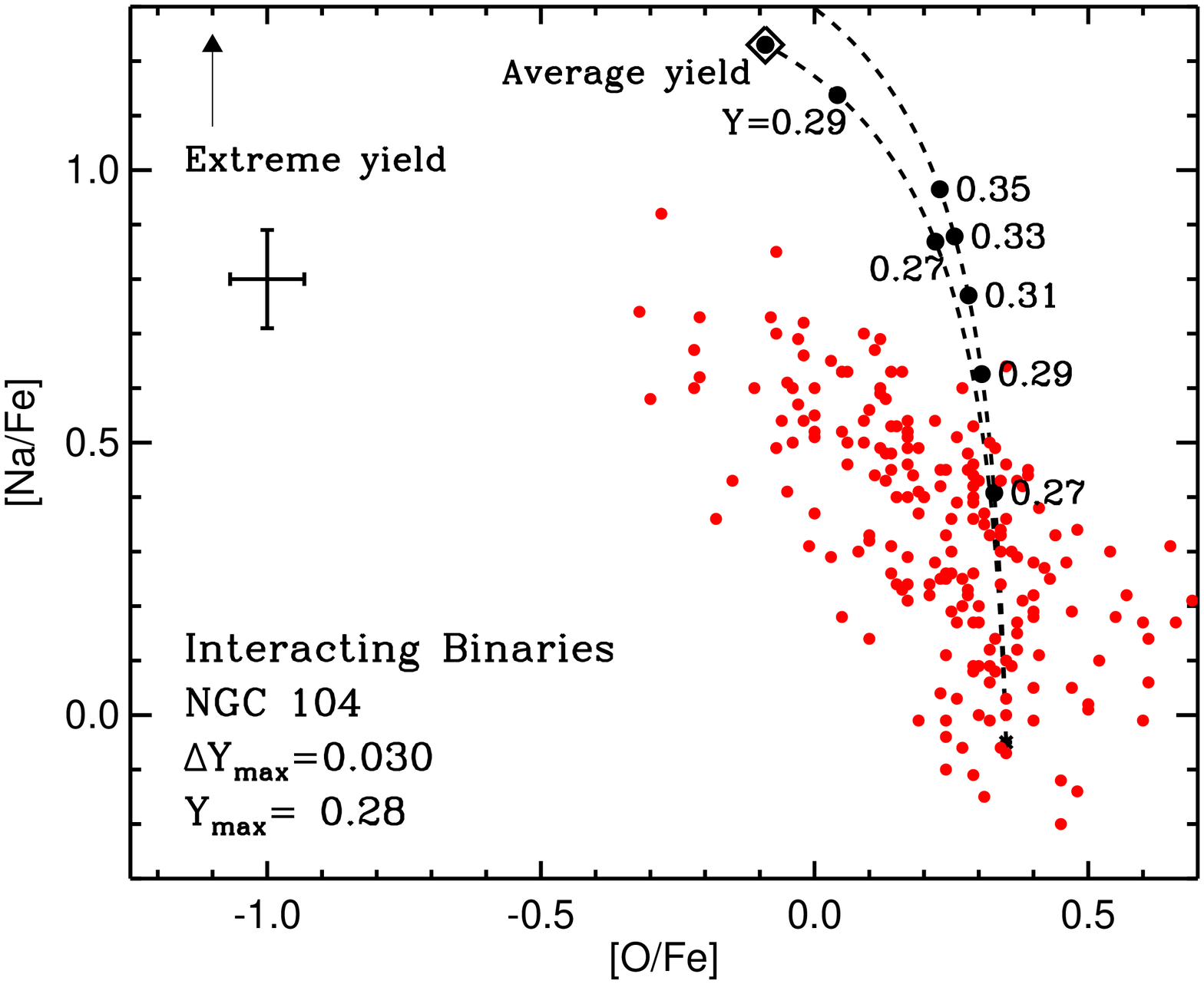}
\caption{Dilution models for NGC~104 (47~Tuc) for four of the scenarios (and yields) considered in the present work.  For the upper two panels the region allowed by the constraints set by the estimated maximum He spread is shaded in grey.  {\bf Top left:} Dilution model for the AGB scenario based on the yields from D'Ercole et al.~(2010).  The yields from the AGB stars are shown as solid lines, while the dashed lines show the resulting abundances when mixing the AGB ejecta with primordial material for different AGB masses and mixing fractions (f$_{\rm mix}$ - see \S~\ref{sec:models}).  Dash-dotted lines connect dilution models with the same He abundance.  {\bf Top right}:  Similar to the previous panel but now using the yields from Ventura et al.~(2013).  {\bf Bottom left:} Dilution model for the FRMS scenario.  Two yields are used (open diamonds - f$_{\rm mix}=1$) which represent the IMF weighted yields for two different assumptions about the form of the stellar IMF.  Dashed lines show the dilution models for these yields and the points mark specific He abundances. {\bf Bottom right:}  Similar to the bottom left panel, the dilution model for the interacting binary scenario (de Mink et al.~2009).  Two yields are used (open diamonds, although one is out of the plotting range), the first is for an average yield, while the second is for the most extreme yield from the model.  None of the models are able to successfully reproduce the observed range in Na and O, and at the same time match the constraints posed by He.  Note the different axes scales for the top and bottom right panels.}
\label{fig:n104}
\end{figure*}

In the upper left panel of Fig.~\ref{fig:n104} we show the yields and dilution models adopting the values of D10.  For comparison we also show the observed spread of stars in NGC~104 (47~Tuc) as filled (red) circles.  Additionally, NGC~104 has been studied photometrically by di Criscienzo et al. (2010) and Milone et al. (2012c) who found a maximum He spread ($\Delta(Y)_{\rm max}$) of 0.03.  If these yields (D10) are adopted, the region permitted in Na--O space, based on the He constraints, is shown as a shaded region.  Note that cluster stars are found with O abundances significantly below the allowed range (i.e., shaded region), which would require $\Delta(Y)\sim0.1$ in order to explain them with this model/yields.  Hence, it is clear that the observations of NGC~104 cannot be reproduced by the AGB yields of D10.

\begin{figure*}
\centering
\includegraphics[width=16cm]{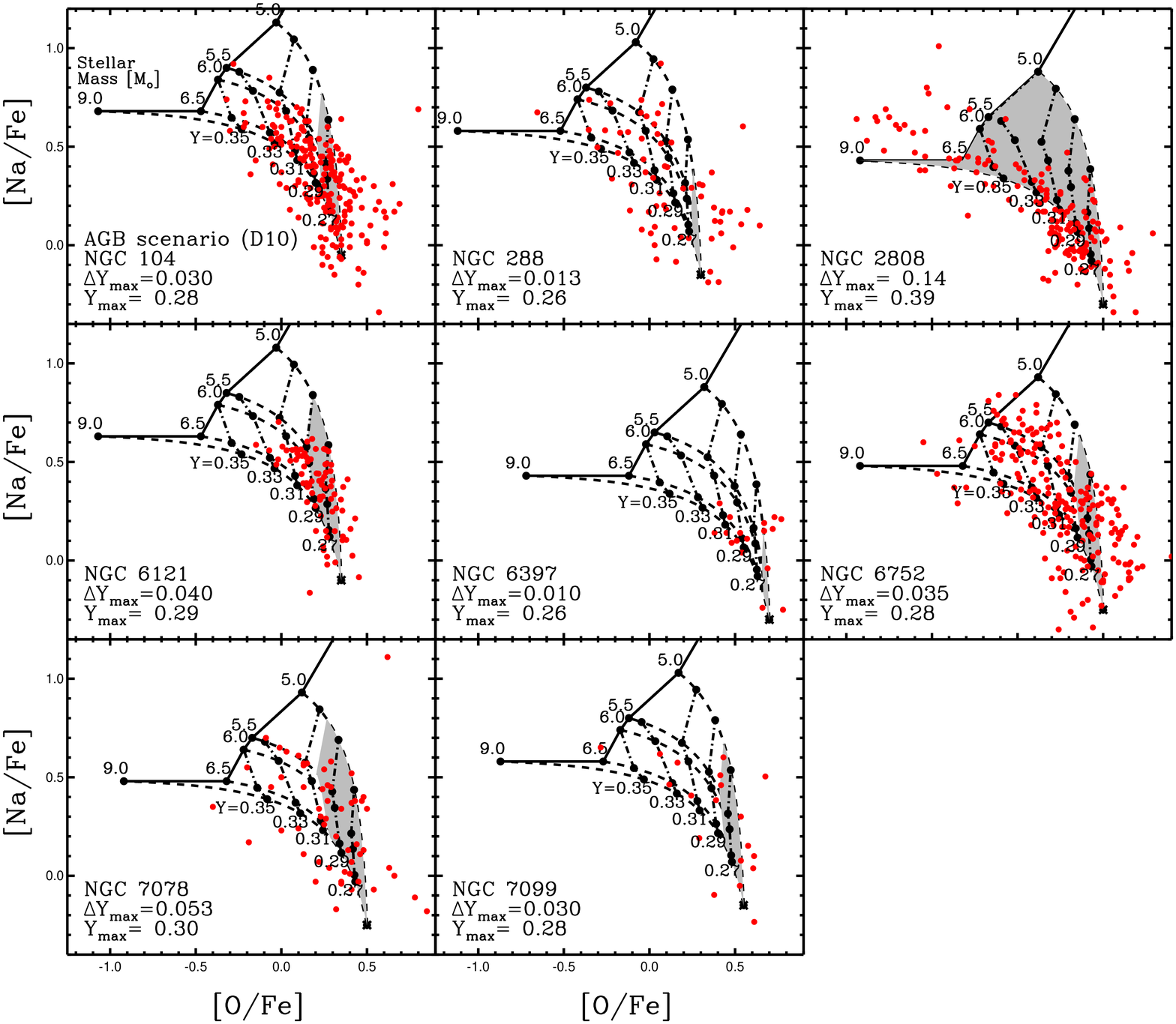}
\caption{Dilution models for a sample of GCs, based on the AGB scenario and using the yields of D'Ercole et al.~(2010).  See Fig.~\ref{fig:n104} for typical errors.  Only stars with measurements of [Na/fe] and [O/Fe] are used, i.e. upper limits are not included.}
\label{fig:agb_d10_tot}
\end{figure*}

In Fig.~\ref{fig:agb_d10_tot} we show similar models for eight other clusters with available data on the Na--O spreads as well as He spreads.  We note that all clusters have stars that extend much further in Na--O space than would be expected given the observed constraints on the He spreads.  As has been noted elsewhere (e.g. Doherty et al.~2014) AGB stars are not able to reproduce the extreme stars in NGC~2808.  

D'Ercole et al.~(2012) have used these same yields and a similar dilution model (albeit with substantially more free parameters as they attempt to match relative numbers of stars of differing compositions, hence timing of the pollution/dillution is important, although the allowed range in Na, O and He is the same as found in the present work) and have applied it to the observations of Marino et al.~(2008) of NGC~6121 (M4).  The authors find a good match between their model and the observations, simultaneously reproducing the Na, O and He spreads, although they predict a relatively small population of extreme He enriched stars that are not found in the photometric analysis.  However, the observations of Carreta et al.~(2009), used in the current work, contain a number of stars with lower O abundances than seen in the Marino et al.~(2008) catalogue.  These stars require significant He enrichment for the D10 yields, which explains the differences between the results presented here and those of D'Ercole et al.~(2012).  However, we note that NGC~6121 and NGC~7078 are the two clusters with the smallest inconsistencies between observations and model predictions.

Based on the observed spreads in Na--O in most clusters in the present sample, He spreads of $\Delta(Y)=0.1$ should be the norm.  However, as seen in the compilation of Milone et al.~(2014), more typical spreads are $\Delta(Y)=0.01-0.05$.

We have also investigated the yields of a mixture of AGB stars of differing masses.  This is meant to represent a situation where the mass reservoir within a cluster builds up for an extended period, before the second generation of stars forms, so that the second generation is made up pristine material as well as AGB stars of differing masses.  In order to see the maximum affect we took the D10 yields for $9~\msun$ and $5~\msun$ stars.  These were chosen in order to obtain the largest range in [O/Fe] as well as to minimise the spread in He (the $5$~\msun\ models produce the lowest amount of He).  However, the He yields for AGB stars do not vary strongly as a function of mass, so even the lowest mass stars still contribute significant amounts of He ($\Delta Y=0.07$).  Because of this the resulting dilution diagram does not change significantly, in terms of the He production, than for dilution models considering a single stellar mass (i.e., those shown in Fig.~\ref{fig:agb_d10_tot}).  Hence, the problem discussed above, of AGB stars producing too much He in order to fit most of the observed clusters still remains even if polluter stars of different masses are considered.

\subsubsection{Yields from Ventura et al.~2013}

\begin{figure*}
\centering
\includegraphics[width=16cm]{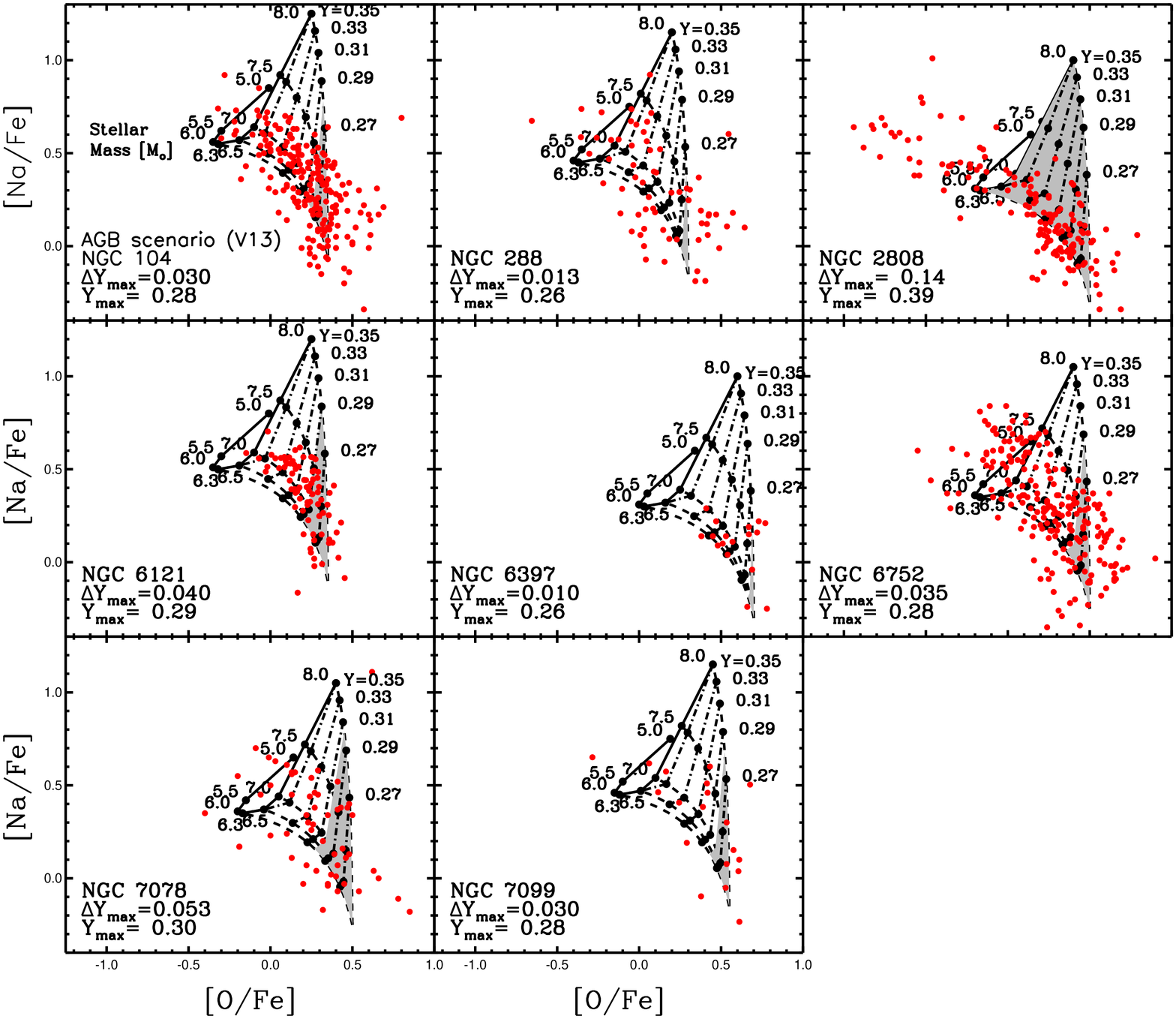}
\caption{Dilution models for a sample of GCs, based on the AGB scenario and using the yields of Ventura et al.~(2013).  See Fig.~\ref{fig:n104} for typical errors.}
\label{fig:agb_v13_tot}
\end{figure*}

In the upper right panel of Fig.~\ref{fig:n104} we show the yields and dilution models adopting the yields of V13.  These yields come directly from AGB evolution calculations, and are not adjusted to match the patterns observed in GCs (contrary to the yields of D10).  For the present analysis we adopt the yields for $Z=10^{-3}$, but models with higher and lower metallciity are discussed in \S~\ref{sec:metallicity}.  These yields suffer from the same problems as the D10 yields, which were discussed above.  Again, the observed stars display much large spreads in O than can be accommodated by the model and the constraints imposed by the observed small He spreads.  In Fig.~\ref{fig:agb_v13_tot} we show dilution models based on the V13 yields for NGC~104 and seven other GCs.

\subsubsection{Yields from Doherty et al.~2014}

As discussed in \S~\ref{sec:models}, the yields from Doherty et al.~(2014) have Na abundances significantly lower than V13, but have similar He abundances.  Hence, as noted by these authors, the super-AGB stars considered in that work, cannot be responsible for the abundance spreads seen in GCs.

\subsubsection{Yields combing different masses}

In the comparisons done so far, between the expected yields plus dilution and observations, we have taken the ejecta of a single AGB star and diluted it with pristine material.  The results show that such combinations cannot reproduce the observations, as the resultant material (i.e., the ``second generation" stars formed) would be expected to have He spreads much larger than that found in the observations.  One could imagine, however, that different masses of AGB stars contribute to the gas reservoir that is then mixed with primordial material.  This is what is assumed in the FRMS scenario, discussed in \S~\ref{sec:frms_scenario}.

To see if this mixture can alleviate the problem with the AGB yields discussed above, we maximise the effect by combining the ejecta of a 9~\msun\ AGB star, a 5~\msun\ AGB star and the primordial material. This is a maximum effect as the 5~\msun\ AGB star produces the least amount of He in all AGB models in the literature (e.g., D10; V13; Doherty et al.~2014).  For AGB yields for lower mass stars the C+N+O yields will no longer be constant, in conflict with the observations (see D10). Additionally, the yields of lower mass stars would also not sufficiently deplete O, also in conflict with the observations.

\begin{figure}
\centering
\includegraphics[width=8cm]{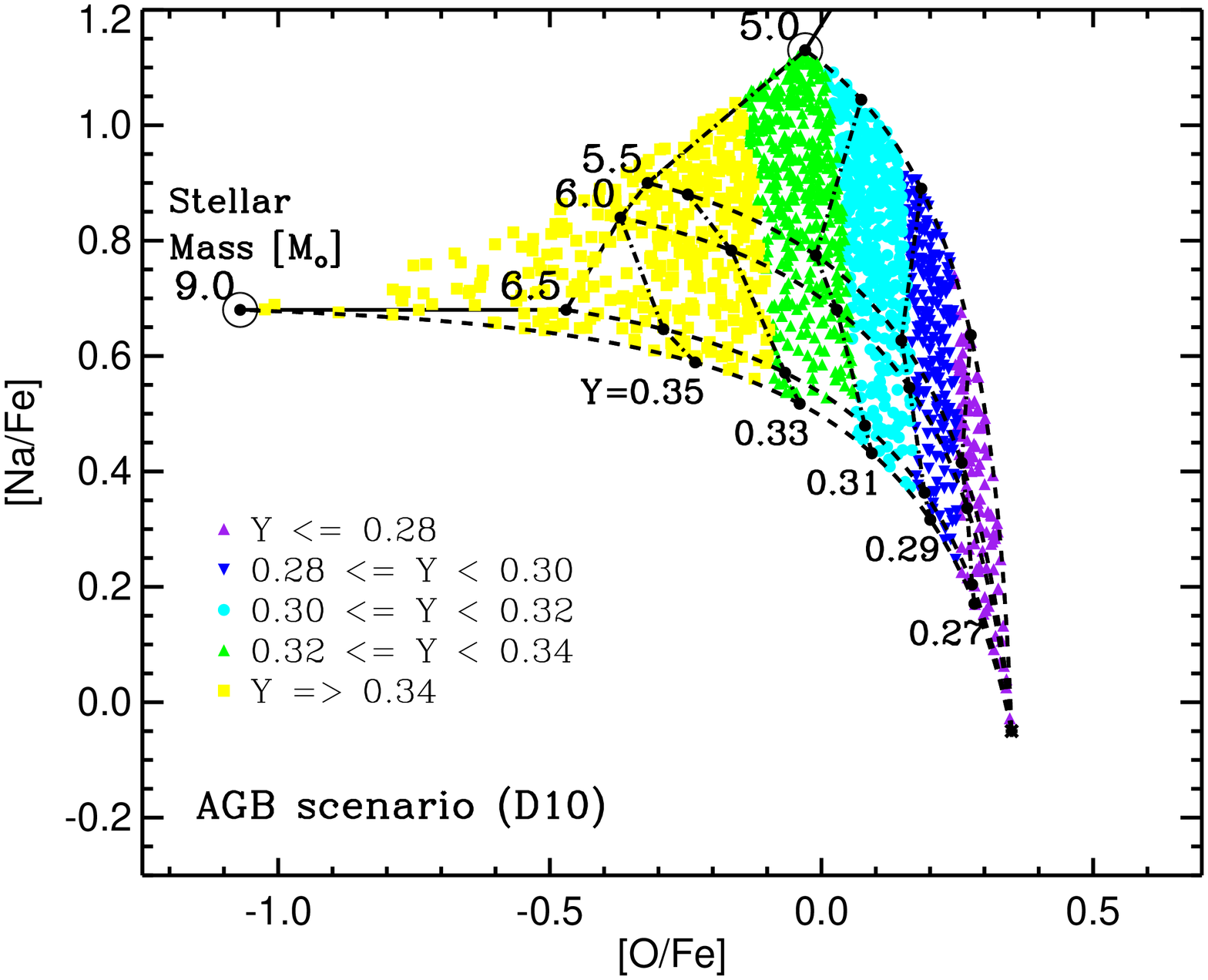}
\caption{Similar to the top left panel of Fig.~\ref{fig:n104}, the dilution model for the D10 yields is shown, however in this case we use three sources of material, the ejecta of a 9~\msun\ AGB star, a 5~\msun\ AGB star, and primordial material.  This maximises the effect of the resulting He spread, as the yield from the 5~\msun\ AGB has the smallest He yield, while the 9~\msun\ model has the largest. The grid is the same as shown in the top left panel of Fig.~\ref{fig:n104}. For a combination of these three yields, the result He abundance is shown in different colours/symbols for differing amounts.}
\label{fig:n104a_combo}
\end{figure}

As our test case we use the D10 yields, and create 1500 dilution models with differing combinations of the amount of material from the three different sources.  The results are shown in Fig.~\ref{fig:n104a_combo}, where the grid is the same as in top left panel of Fig.~\ref{fig:n104}.  The different colours/symbols represent mixtures of different He contents.  The basic result is that the level of He enrichment does not change significantly by including combinations of the two extreme AGB yields (differing in mass).  For a given abundance of Na and O, the expected He is similar to the 'simple' dilution case of a single AGB ejecta yield plus primordial material.  The reason for this is simple, that all AGB ejecta that significantly deplete O all produce large amounts of He, so the expected changes in the He enrichment are very similar (see Fig.~\ref{fig:metallicity}).

\subsection{FRMS scenario}
\label{sec:frms_scenario}

In the FRMS scenario, a second generation of stars forms out of a combination of material ejecta from rapidly rotating massive stars and material left over from the formation of the first generation.  For a more complete summary of the full scenario, see Krause et al.~(2013) and Bastian et al.~(2014).  The FRMS ejecta is rich in He and Na, and depleted in O (e.g., Decressin et al.~2007).  In the bottom left panel of Fig.~\ref{fig:n104} we show the yields (open diamonds) and dilution models from Decressin et al.~(2007).  As can be seen, the choice of the form of the IMF does not strongly affect the resulting yields and models.  As was found for the AGB scenario above, in order to reproduce the observed range in Na and O, the models predict large spreads in He, $\Delta(Y)_{\rm predicted}=0.11$, which are inconsistent with the observed constraints $\Delta(Y)_{\rm observed}=0.03$.

\begin{figure*}
\centering
\includegraphics[width=16cm]{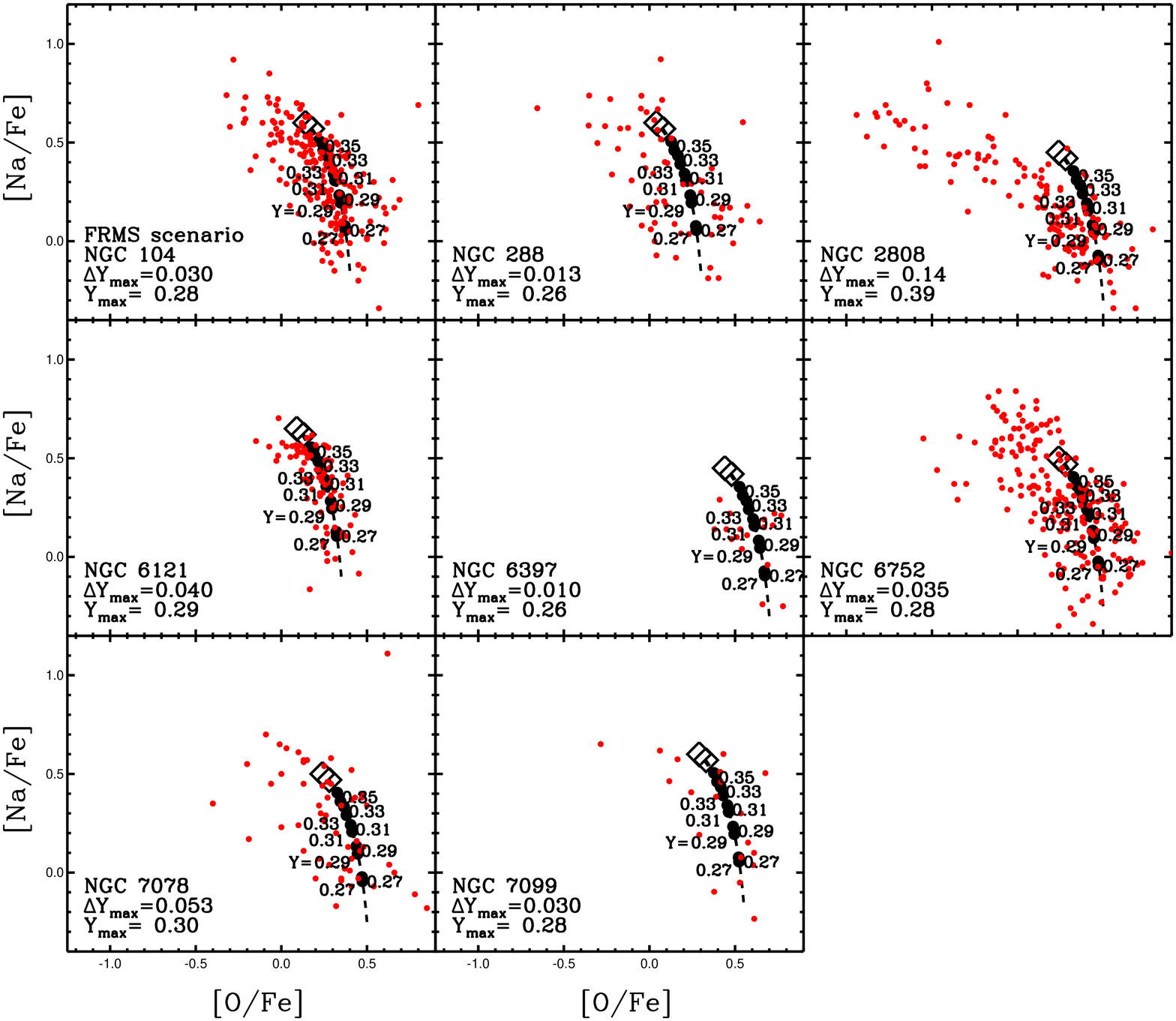}
\caption{Dilution models for a sample of GCs, based on the FRMS scenario and using the yields of Decressin et al.~(2007).  Open diamonds denote the expected yields (IMF weighted) for two different assumptions about the slope of the IMF.}
\label{fig:frms_tot}
\end{figure*}

This problem can be seen more generally for all clusters in our sample in Fig.~\ref{fig:frms_tot}.

For the yields of the FRMS scenario, we have adopted the time averaged yields from the massive stars, i.e. that all the material is collected from the massive star before being diluted and forming new stars.  Decressin et al.~(2007) have shown that such yields do not produce enough variation in O and Na to match the observations of clusters like NGC~6752.  To get around this problem, the authors suggest that the second generation stars will form as the material is being ejected from the high mass stars, hence the final abundance of the formed stars will depend on when the material is ejected (as the abundances of the ejected material change over time) and how much dilution happens to that material.  By adjusting these parameters, large ranges of O and Na are possible in second generation stars.  However, in such a case the Na-rich, O-poor second generation stars would be expected to also be very rich in He (up to $Y=0.7$), which is even more problematic than the time-averaged yields. 

\subsection{Massive interacting binaries}

Interacting massive star binaries are the next source of the enriching material that we consider in the current work.  These sources were adopted in the ``Early Disc Accretion" (EDA) scenario of Bastian et al.~(2013b), where low mass pre-main sequence stars with proto-planetary discs sweep up material processed by the interacting massive star binaries, which eventually gets accreted onto the host low-mass star.  One caveat to the yields adopted here is that they are from a single interacting binary system (dM09).  Hence, it is currently unclear whether the published yields are representative of typical interacting binaries.

A comparison between the expected yields and simple dilution models with the observations is shown in the lower right panel of Fig.~\ref{fig:n104}.  It is clear that the extreme yields are not consistent with the observed small He spreads observed within the cluster.  However, in this particular case, the average yields do not strongly violate the constraints imposed by the He measurements.  Nevertheless, it is also clear that the models do not reproduce the shape of the Na--O correlation in detail, with the models predicting less O depletion than observed.

\begin{figure*}
\centering
\includegraphics[width=16cm]{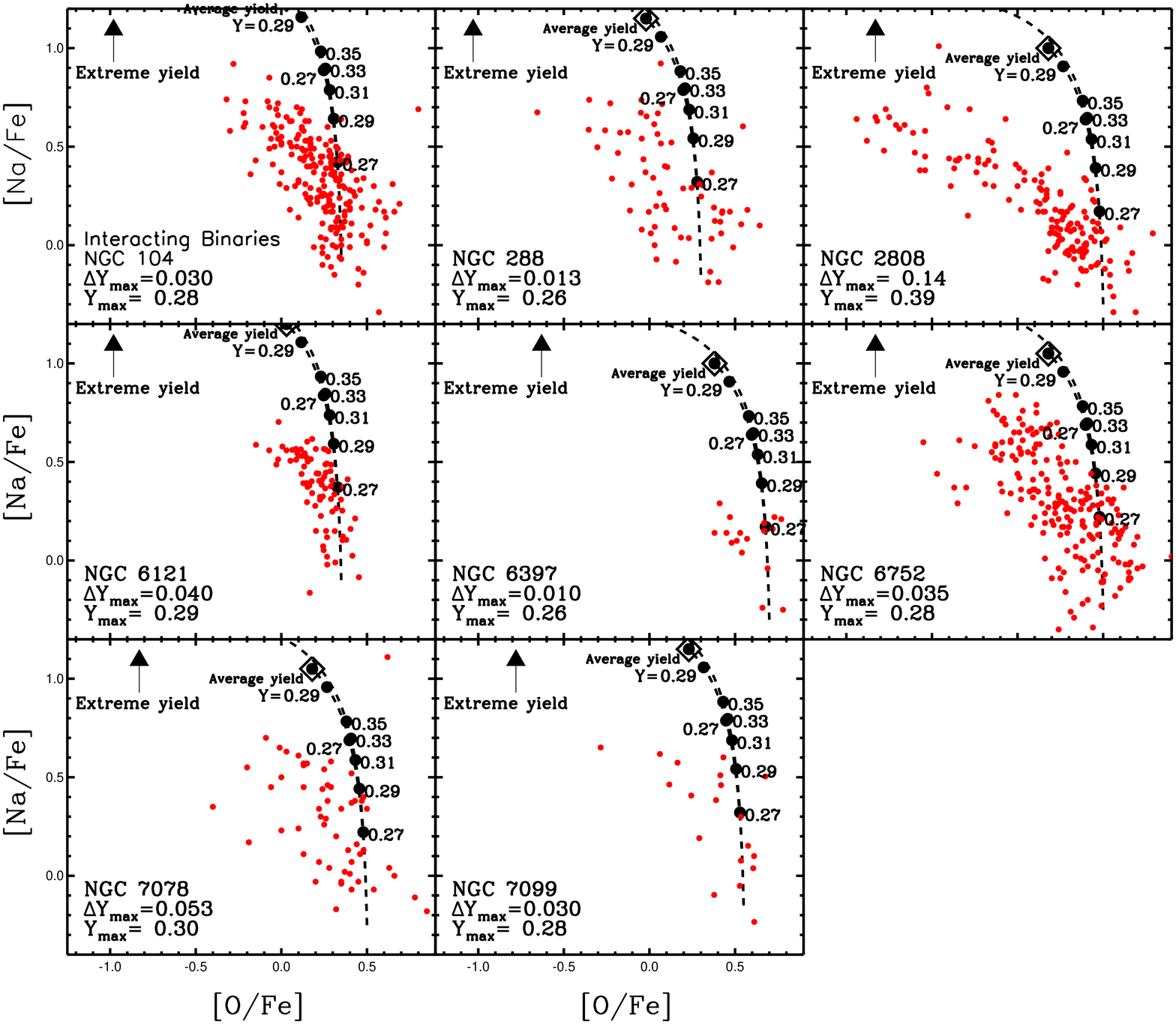}
\caption{Dilution models for a sample of GCs, based on the Interacting Binaries scenario and using the yields of de Mink et al.~(2009).  Two models are shown, one for the 'average yield' (open diamond) and another for the 'extreme yield' (i.e., the most extreme abundances reached in the ejected material) which is off the plotting range ([O/Fe]$_{ex} =-1.13$, [Na/Fe]$_{ex}=1.3$).}
\label{fig:ib_dm09_tot}
\end{figure*}

A comparison between the yields and dilution models for all the clusters in our sample is shown in Fig.~\ref{fig:ib_dm09_tot}.  We find similar results as was found for NGC~104.  Due to the lower He enrichment by the average yields, the He constraints are not as strongly violated as they are for the other types of polluting stars.

As noted in \S~\ref{sec:models} and discussed in more detail below, one of the largest strengths of the massive interacting binary model is in the potential stochasticity inherent in the resulting yields (discussed further in \S~\ref{sec:disc1}).  However, it is not clear if IBs could produce the observed ranges of Na and O, while also producing strongly varying amounts of He (for a given range of Na and O).  Hence, further models are required to test the general validity of these types of stars as potential sources of pollution within GCs.

\subsection{Very massive stars}
\label{sec:vms}

The final source for the enriching material that we consider here is that of Very Massive stars (VMS), i.e. stars with masses exceeding $10^4$~\msun.  These stars were suggested by Denissenkov \& Hartwick~(2014) as potential polluter stars, as their yields generally match that observed in GCs (i.e. deficient in O, enhanced in Na and He).  While this potential source has not been developed into a full scenario, i.e. how the processed material from these very massive stars finds its way onto stars within the cluster (or how it forms new stars), we can test general predictions of the model.  Additionally, we note that the material lost by the VMS is assumed to be ejecta after the core He changes by $\Delta Y = 0.15$ in order to match the abundances of NGC~2808.  If the star is left to continue burning H, the He fraction will increase substantially, compounding the problem of the overproduction of He, discussed above for other models, and in more detail in \S~\ref{sec:disc1}.

\begin{figure*}
\centering
\includegraphics[width=16cm]{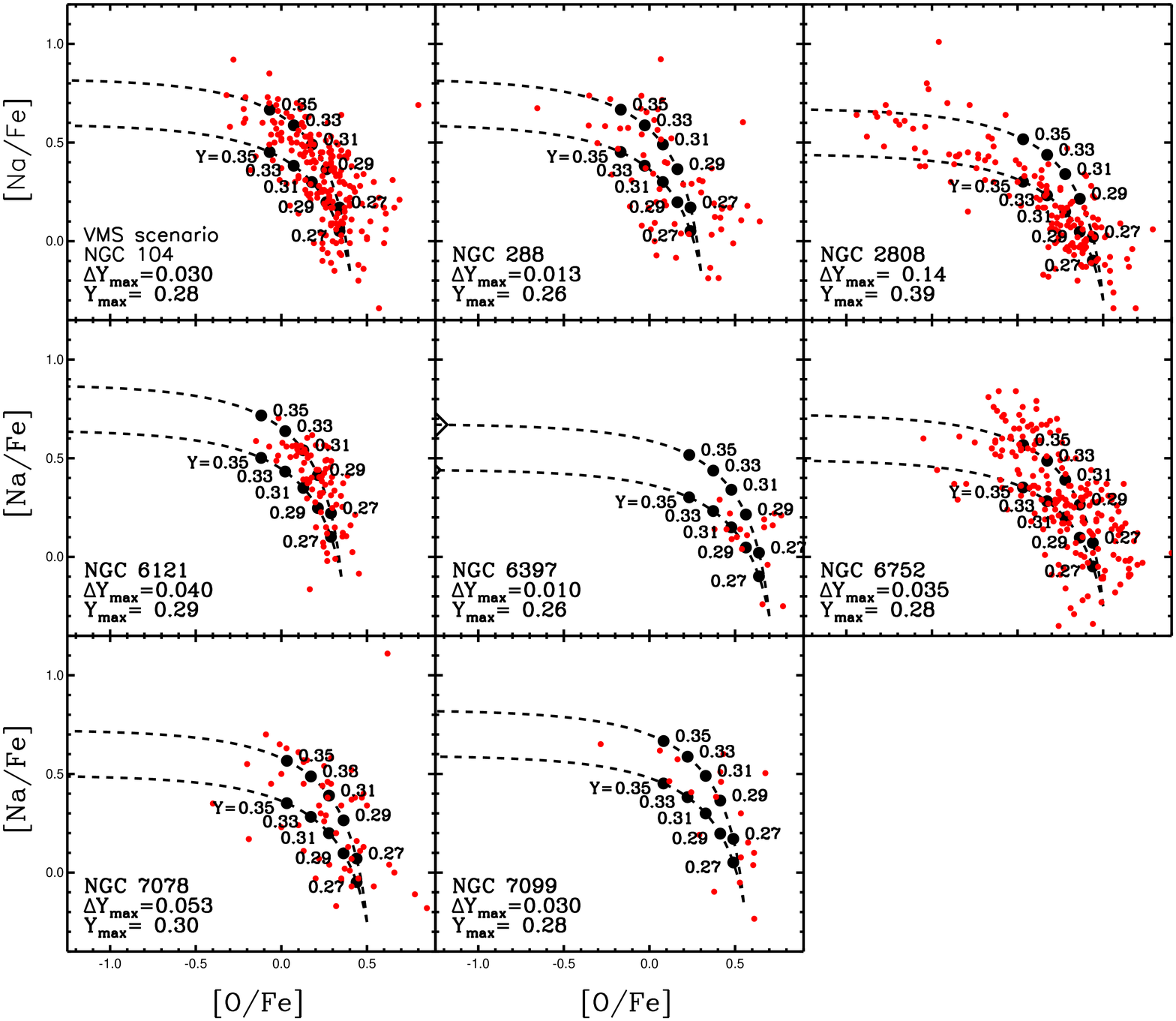}
\caption{Dilution models for a sample of GCs, based on the Very Massive Star scenario and using the yields of Denissenkov \& Hartwick~(2014).  Two models are shown, one for a star of $5\times10^4$~\msun (top dashed line) and $7\times10^4$~\msun (lower dashed line). }
\label{fig:vms_tot}
\end{figure*}

We show dilution models using two VMSs, with masses of $4\times10^4$ and $7\times10^4$~\msun, compared to observations of eight GCs in Fig.~\ref{fig:vms_tot}.  As was seen with the other sources that use high mass stars, the VMS scenario generally produces too much He for a given range in Na--O.  While the formation of such VMSs may be stochastic in nature, the resulting yields due not appear to possess the stochasticity required to explain the wide range of observed GC properties.

\section{Empirical Dilution Models}
\label{sec:empirical}

In addition to the physically motivated polluter sources discussed above, we have also tested empirical dilution models.  For this, we adopt the ``extreme abundances" of enriched stars in clusters, assign the maximum He values to these stars, and then use this material as the polluter material in the dilution models.  This is similar to that done by Carretta~(2014), who studied NGC~2808, and found that the ``intermediate" enriched stars could not be reproduced by a combination of the extreme stars in the cluster and primordial material, suggesting the different sources are responsible for the pollution of the intermediate and extreme enriched stars.

\begin{figure}
\centering
\includegraphics[width=8cm]{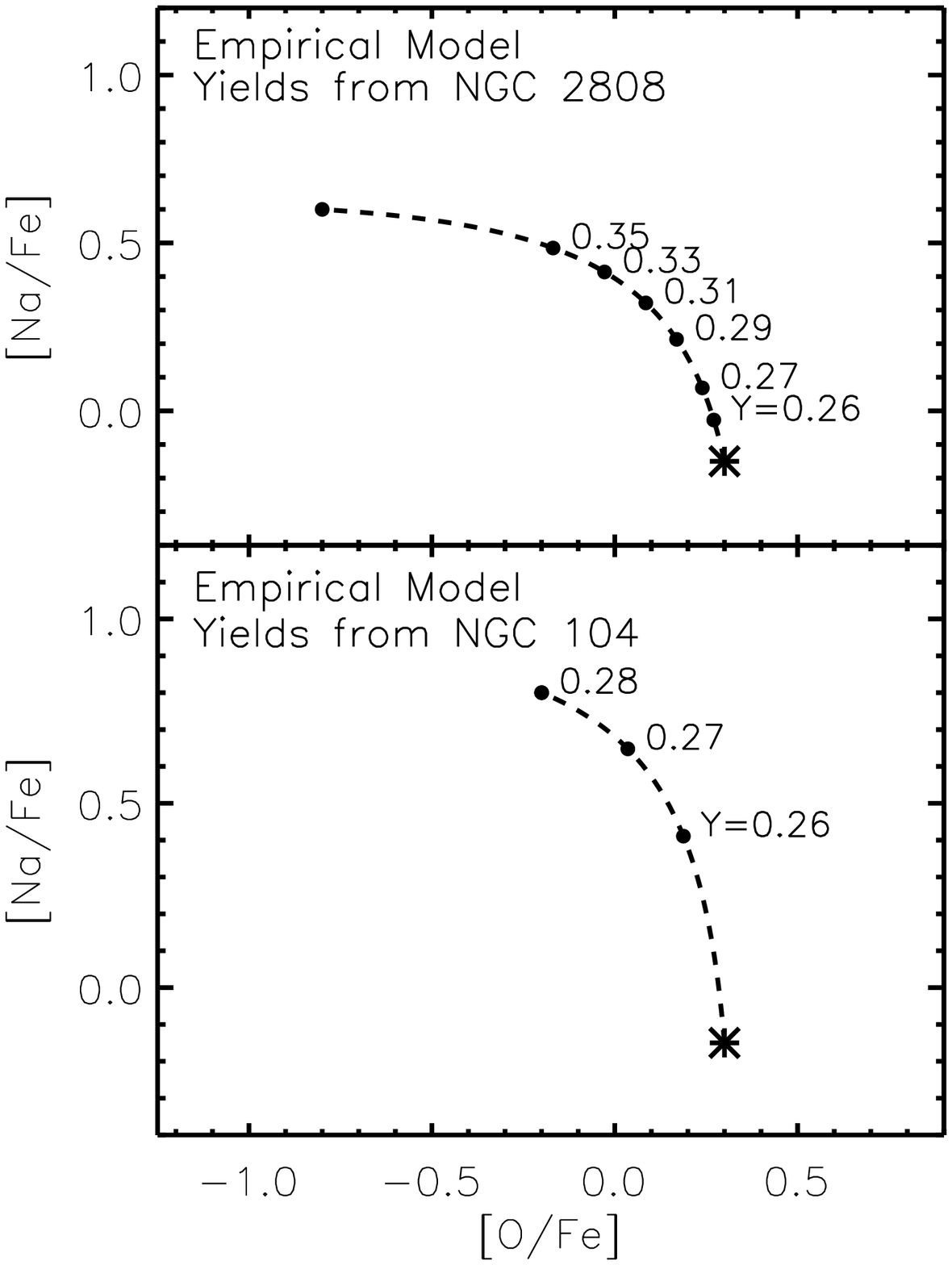}
\caption{Empirical dilution models using as yields the extreme stars of NGC~2808 (top panel) and NGC~104 (bottom panel).  For the yields, we adopted the values of [Na/Fe] and [O/Fe] of the most extreme star in the cluster, and we also assigned the maximum He spread of the cluster to this point.}
\label{fig:empirical}
\end{figure}

The goal of this test is to see if any generalised polluter model could work, which if so, could offer a clue as to what source is responsible for the enriched material.  To begin, we use the extreme stars of NGC~2808.  The results are shown in the top panel of Fig.~\ref{fig:empirical}.  Due to the large He spread (the inferred maximum He value in NGC~2808 is Y$_{\rm max}=0.39$), the resulting dilution models also have high He values, that are incompatible with the majority of the clusters discussed in the present work.  This model behaves in a very similar way as the yields of D10, which is not surprising, given that D10 modified the AGB yields in order to match this cluster.  However, as was the case for the D10 yields, these empirical yields cannot explain the other clusters, so whatever source was responsible for the enrichment of NGC~2808, can not be the source for the other clusters discussed here.

Additionally, we use the extreme stars of NGC~104 to make an empirical dilution model for this cluster.  The results are shown in the bottom of Fig.~\ref{fig:empirical}.  Due to the lower He spread in this cluster (the inferred maximum He value in NGC~104 is 0.28), the resulting model may, in principle explain some of the GCs in our sample (e.g., NGC~6121, 6397, NGC 6752).  However, unsurprisingly, the model cannot explain NGC~2808, nor any cluster with He spreads larger than $0.03$ (e.g., NGC~7078).  It also over predicts the amount of He in cluster like NGC~288, which have similar Na--O spreads as NGC~104, but have much lower He spreads.

We conclude that no single polluter source can explain all the GCs in the current sample.  As was found with the more physically motivated model yields (AGBs, FRMS, IBs, VMS), due to the range of observed properties, and models that can explain clusters like NGC~2808, will necessarily fail when matching clusters like NGC~104 and 288 (and vice versa).  However, even models with non-extreme GCs like NGC~104 for the empirical yields, while providing a closer fit for most clusters in our sample, are not consistent with GCs with either very low He spreads (NGC~288) or relatively high He spreads (NGC~7078).

Hence, no single polluter model, regardless of the uncertainties in the yields, can explain the observed abundance trends.  Whatever mechanism is responsible for the abundance spreads in GCs, it must have a large stochastic component to it, producing some GCs with large He spreads and others with small He spreads, but at the same time producing similar Na--O spreads between clusters with very different He spreads.

\section{Discussion}
\label{sec:discussion}

\subsection{Constraints from the Chemical Analysis}
\label{sec:disc1}
As shown above, the general result is that {\em all} the proposed polluters produce too much He in order to fit the observed range in Na and O for most clusters in the present sample.  In particular, the models predict that any significant range in O variations must be accompanied by a large He spread as well.  Such large He spreads are not consistent with observed colour magnitude diagrams of many the GCs studied in the present work (e.g., Milone et al.~2014).  Some GCs do display significant He variations (e.g., NGC~2808; Piotto et al.~2007).  However, others with similar (although slightly less extended) Na--O spreads host He spreads of $\Delta(Y)$ of 0.013-0.035, instead of the expected $\Delta(Y)$ of 0.1-0.15.

It appears to be a general problem that the enrichment sources share, that they predict too much He for a given amount of Na--O variation, except in the most extreme (e.g., NGC~2808) cases.  Of the five model yields discussed in the present work, the interacting binary model does the best in describing the abundance variations observed in the majority of clusters, if the average yields are adopted, due to the fact that it has the lowest He yield of any of the four models.  However, if the extreme yields are adopted, the model fails to reproduce the observed trends in many of the clusters (e.g., NGC~104; NGC 288, \& NGC 6752).

Hence, it appears to be a generic problem for all enrichment scenarios, that similar spreads in some elements (Na--O) are not accompanied by the same spreads in He.  While there are uncertainties in the yields for all  sources of the enriching material, the key point here is that if a model matches a cluster like NGC~2808 with large He spreads, it will, by construction, not fit the other clusters with large Na and O spreads but with small He spreads.    Since the majority of clusters studied to date have relatively small He variations ($\Delta(Y) \lesssim 0.05$) enrichment sources with large He yields (for a given amount of Na--O variation) are disfavoured.  Models with modest He yields but large Na--O variations are more generally applicable for the average cluster, although they will necessarily miss the more extreme He-enriched clusters.

It appears that only models with an intrinsically large stochastic component can potentially fit the array of observed properties in GCs.  While all scenarios can vary the fraction between the amount of pure ejecta (AGB or massive stars) and primordial material that goes into forming later generations of stars, the results presented here show that this is not sufficient to reproduce the observed trends.  The reason is that the dilution of the ejecta will change all elements together, resulting in direct predictions of how each element varies with one another.  Such predictions are in contradiction with the observed abundance patterns.

For most globular clusters, with current or initial masses in excess of a few times $10^5$~\msun, their stellar IMFs should have been fully sampled.  Even for the FRMS scenario, which uses stars more massive than $20$~\msun, the IMF is expected to be fully sampled for typical Chabrier-type IMFs, hence the models do not expect much stochasticity in the ejecta yields between clusters.  This is even more true for the AGB scenario, which uses stars with masses of $\sim5-9$~\msun.  So it is unlikely that these models will be able to provide the stochastic element required by the observations.  Very massive stars may be stochastic in their formation, however their yields do not appear to have the necessary stochastic component to them. Interacting binaries can potentially provide the necessary stochasticity, as the yields are dependent on the 1) mass ratio of the interacting stars, 2) the mass of the primary star, and 3) the evolutionary state of both stars when the interaction takes place.  However, to date, only a single model for the yields of interacting binaries has been published, hence it is too early to know if interacting binaries actually can provide the necessary stochasticity.


Cassisi \& Salaris~(2014) and Salaris \& Cassisi~(2014) have investigated the ``Early Disc Accretion" scenario and the expected yields and abundance trends in detail, using the yields of dM09.  In addition to Na, O and He, these authors also used Li abundances as further constraints.  They found that in NGC~6752 the dM09 yields, along with the EDA scenario (Bastian et al.~2013b) were not able to match the observations.  From the middle right panel of Fig.~\ref{fig:ib_dm09_tot} it is clear why this is the case.  NGC~6752 has a large O spread (and Na) and a small He spread.  Hence, the interacting binary model, along with all scenarios discussed here will not be able to match this cluster.  The authors could place less stringent constraints on NGC~104 (47~Tuc) and NGC~6121 (M4) but it was difficult to match the observations without appealing to large observational errors or fine tuning the model.  Based on the observations present here, we would expect NGC~288 and NGC~7099 to have similar problems.  For NGC~2808 Cassisi \& Salaris~(2014) could fit the observed abundances with the interacting binary yields, but only by adopting the extreme yields, due to the high He yields of those models.

One potential solution to the problem exposed here for the AGB scenario, would be if different GCs were polluted by different mass ranges of AGB stars.  For example, if only massive clusters were able to retain the ejecta of high mass AGB stars that are highly enriched in He (because of the second dredge-up) and Na, and highly depleted in O, while lower mass clusters could only retain the ejecta of lower-mass AGB stars (that do not undergo the second dredge-up, hence are not enhanced in He at the surface).  As a general note, however, by restricting the mass range of AGB stars allowed to contribute to the enrichment, the mass-budget problem will be further aggravated.  Additionally, in the models of AGB yields of V13 and D10, lower mass AGB stars ($<5$~\msun) produce too much O to match observations, and below $\sim5$~\msun\ the sum of [(C+N+O)/H] varies outside the range allowed by observations (e.g., Cohen \& Mel{\'e}ndez~2005). Models that invoke high mass stars ($>20$~\msun) to provide the enrichment are not sensitive to the mass range covered, due to the similarities in the expected yields for a wide range of stellar mass (e.g., Decressin et al.~2007; Denissenkov et al.~2014).  Hence, while potentially feasible, the pollution of clusters by stars in different mass ranges does not appear to solve the general abundance problem discussed in the present work.

\subsection{The Effect of Metallicity}
\label{sec:metallicity}

The GCs discussed in the present work span a relatively large range in metallicity from [Fe/H]$=-0.72$ (NGC~104) to [Fe/H]$=-2.37$ (NGC~7078).  Hence, it is reasonable to ask whether differences in the metallicity of the clusters, which may cause different yields from the polluting stars, may be the cause of the observed differences in He spread and Na--O extent.  Metallicity cannot be the full explanation, however, as the two most different clusters in terms of their He spreads, have very similar metallcities (NGC~288, $\Delta Y=0.013$, [Fe/H]$=-1.32$; NGC~2808, $\Delta Y=0.14$, [Fe/H]$=-1.14$).  Additionally, two clusters with similar He spreads, as well as Na--O extents, have very different metallicities (NGC~104, $\Delta Y=0.03$, [Fe/H]$=-0.72$; NGC~7099, $\Delta Y=0.03$, [Fe/H]$=-2.26$).

\begin{figure}
\centering
\includegraphics[width=8cm]{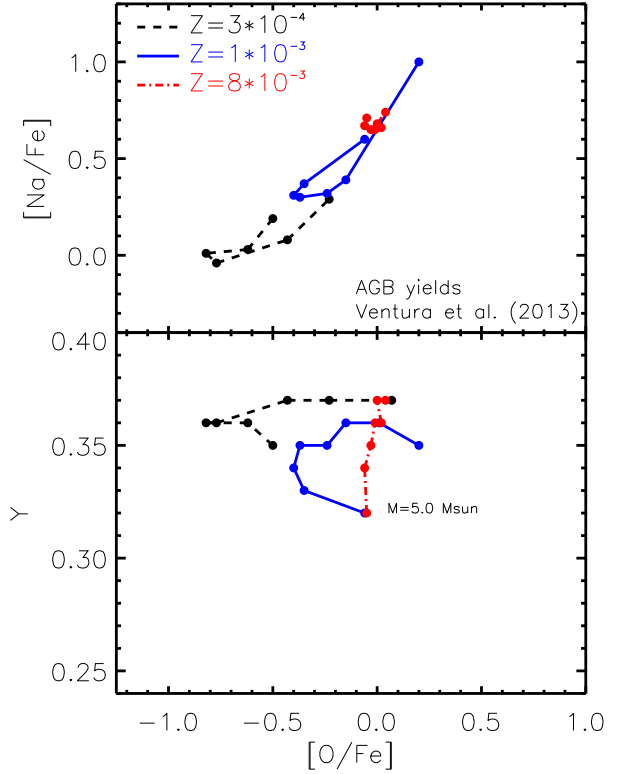}
\caption{The predicted yields of AGB stars of different masses for three different metallicities (given in the top panel).  The top panel shows the [O/Fe]--[Na/Fe] relation, while the bottom panel show the expected He yield as a function of [O/Fe].  Note that the He yield is very similar for all masses and all metallicities.}
\label{fig:metallicity}
\end{figure}

V13 have calculated the expected yields from AGB stars for three different metallcities ($Z=3\times10^{-4}, 1\times 10^{-3}$, \& $8\times10^{-3}$ - corresponding to [Fe/H]$ \approx -1.82, -1.3, \& -0.40$, respectively).  We show the results of their calculations in Fig.~\ref{fig:metallicity} for stars of masses between $5$ and $8~\msun$.  In the lower panel, we show the He yield as a function of [O/Fe].  Despite the large range covered in metallciity, the He yield only changes by a small amount ($\Delta Y = 0.03$ between the lowest mass stars of different metallcities).  Hence, He production is largely independent of metallicity, so metallicity differences between the clusters cannot explain the observed range of He spreads between clusters.

Additionally, we note from the top panel of Fig.~\ref{fig:metallicity} that a prediction from this set of calculations, is that in low metallicity clusters, Na should be much less enriched than in high metallicity clusters.  However, we find no such correlation in the observations (see Fig.~\ref{fig:agb_v13_tot}, for example).

\subsection{Additional Problems for Self-Enrichment Scenarios}

\subsubsection{A Correlation Between [N/Fe] and Cluster Mass}
Schiavon et al.~(2013) found a correlation between [N/Fe] and mass for a sample of 78 GCs in M31, spanning a range in mass of $10^5 - 10^{6.5}$~\msun.  The correlation was such that the most massive clusters in the sample had on average $\sim2.5$ times more N (per Fe) than the least massive clusters in the sample.  Since, all clusters in the sample would be expected to have a fully sampled IMF, in most scenarios, the amount of self-enrichment should be independent of cluster mass (the amount of enriched material should be the same per unit stellar mass).  One way to obtain this result would be if lower mass GCs would be diluted with a higher fraction of primordial material, so that their resulting abundance pattern would be closer to that of the first generation.  However, this is counter-intuitive, as lower mass clusters have shallower potential wells, so they would be expected to accrete and retain less primordial material from their surroundings than higher mass GCs.

The only other way to end up with enhanced enrichment (per unit mass) is if the majority of the enriching material is lost at low cluster masses, and have more enriched material be retained by the higher mass clusters.  This is difficult to realise in most self-enrichment scenarios, as they already have a severe mass budget problem, and assume that all enriched material from a first generation is retained within the cluster and used to form subsequent generations.  In the AGB and FRMS scenarios, for example, $90-95$\% of the 1st generation stars must have been lost after the formation of the 2nd generation, in order to satisfy the observed mass budget constraints. Hence, in these scenarios GCs were 10-20 times more massive at birth than they are currently.  For the lower mass clusters in the Schiavon et al. sample, this number would increase to 25-50 in order to reproduce the Schiavon et al. relation (under the assumption that the highest mass clusters do retain 100\% of the processed material). Such large mass loss rates cannot be a common feature of all GC populations (Larsen et al.~2012; 2014).

\subsubsection{A Correlation Between He Abundance and Cluster Mass}
A similar argument can be made for the He spreads observed in GCs.  Milone et al.~(2014) has found that the maximum He spread ($\Delta(Y)_{\rm max}$) is proportional to the absolute luminosity (hence mass) of the GC.  As with the [N/Fe] correlation with GC mass, this is difficult to reconcile with self-enrichment scenarios.  The only way to do this is if lower mass clusters lose large fractions of any enriched material from ``1st generation" stars, whereas high mass clusters retain larger fractions.  Hence, this also worsens the mass budget problems for the enrichment sources (AGBs, FRMS, IBs, VMS) studied in the present work.

\subsubsection{Expected Correlation between Metallicity and Abundance Trends}
In addition to the above points, scenarios for self-enrichment that invoke multiple epochs of star formation all have a general severe mass budget problem, where there are not enough first generation stars to provide the material required to enrich the observed number of second generation stars.  In order to solve this, such scenarios assume that GCs were much more massive at birth, by factors of $\gtrsim10$ for the AGB and FRMS scenarios, respectively (D08; D'Ercole et al.~2010; Schaerer \& Charbonnel~2011; Conroy~2012).  Such mass loss is not supported by observations of clusters and field populations in dwarf galaxies (Larsen et al.~2012; 2014).  The first generation stars lost in this way must be lost after the second (subsequent) generation(s) form.  Models to explain this strong mass loss need to invoke 1) strong tidal perturbations (i.e. that the clusters form in strong tidal fields) and 2) that the young GCs were strongly mass segregated (e.g., D'Ercole et al.~2008).  However, metal-poor clusters are generally thought to form in low mass dwarf galaxies, which were subsequently accreted onto the Milky Way.  Such dwarf galaxies have significantly weaker tidal fields than the inner regions of the Milky Way (used in the D'Ercole et al.~2008 calculations), so it is difficult to see why such metal poor clusters would lose a large fraction of their initial masses.  At the very least a relation between the fraction of enriched stars and cluster metallicity would be expected in such a scenario, which is at odds with observations (Carretta et al.~2010). 

\subsubsection{An Unexpected Inversion in Radial Profiles}

Recently, Larsen et al.~(2015) have studied the radial profiles of the primordial and enriched populations within the Galactic GC M15, based on a combination of ground based photometry in the outer regions and HST imaging of the central region.  Previous work had shown that in the outer regions, the enriched stars were more centrally concentrated than the primordial stars (Lardo et al.~2011).  However, Larsen et al.~(2015) found that this trend reverses in the inner $\sim3$~pc, where the primordial stars are the most centrally concentrated.   The authors investigate the possible role of mass segregation (since He enriched stars have lower masses on the RGB than He normal stars for globular cluster ages) and conclude that the required He spread ($\Delta Y \gtrsim 0.15$) is not compatible with the observed constraints on any spread in He ($\Delta Y \lesssim 0.03$).  

The authors conclude that mass segregation is unlikely to be the cause of the observed inversion, and that the observations of M15 are in conflict with the predictions of all self-enrichment scenarios put forward so far.  How common such inversions are is still an open question (see Larsen et al.~2015 for an in depth discussion of results in the literature).


\section{Conclusions}
\label{sec:conclusions}

We have presented dilution models based on the yields of commonly advocated sources for the enrichment of globular cluster stars; namely AGBs, Fast Rotating Massive stars, Interacting High Mass Binary stars, and Very Massive stars. This study has taken advantage of the recent measurements of the He spread within GCs, combining this with the more traditional measurements of Na and O spreads.  While each type of polluter star can explain some of the observations (e.g., NGC~2808 with its high He abundances and relatively large range in Na and O), all the models fail in describing more typical clusters (e.g., NGC~104, 288, 6752).  The basic problem is that all of the polluting sources produce too much He for a given amount of change in Na and O.  Any of the models that can explain clusters like NGC~2808 will necessarily fail in reproducing clusters like NGC~104 (47~Tuc).

In order to consistently explain all clusters, a high degree of stochasticity is required.  We have produced empirical dilution models based on the ``extreme stars" observed in clusters like NGC~2808 (large He spreads) and NGC~104 (modest/typical He spreads), and found that no single model could explain all of the observed GCs.  Whatever the origin of the abundance spreads, the process must result in a wide variety of chemical trends. In particular, large variations in the He yields for a given change in Na and O are required.  Of the four physically motivated sources of enrichment considered here, only the interacting massive binary model potentially satisfies this criteria (although it is unclear if this model can result in large He differences for a given spread in Na and O).  However, this model has other problems with regards to the observed chemistry (e.g., Lithium - Salaris \& Cassisi~2014).

The current observations suggest that the range of Na and O spreads are not directly correlated with the spread of He within GCs.  Any self-enrichment scenario that invokes material processed by either high-mass stars or AGB stars will necessarily fail on this point (regardless of the uncertainties in the yields of the polluter stars), unless external agents are brought into the model (e.g., some unspecified process that removes He from the ejecta while leaving other elements untouched).  For the high-mass stars, this is due to basic nuclear burning, where the Na enhancement and O depletion takes place at the same location as He production.  For AGB stars this is due to basic AGB evolution, i.e. the hot bottom burning and the second dredge up.

We stress that we have not tested the full scenarios for each model, instead we have given each model the benefit of the doubt regarding their underlying assumptions (e.g., that large reservoirs of gas with the primordial abundance patterns are available to be used at the specific times required).  Rather, we have focussed on the basic yields that are expected for each type of polluter star invoked.  While there are uncertainties in the yields for each polluter type (i.e., AGBs, FRMS, IBs, VMSs), our results show that none are consistent with the range of observed abundance patterns/correlations, and that adjustments to the yields are unlikely to solve the problem.  Instead, the problem appears to be a general one for all self-enrichment scenarios.

The results presented here add to a growing list of problems faced by any self-enrichment scenario.  In particular, the observed trends of [N/Fe] abundance and He spreads with cluster mass are impossible to reproduce unless 1) strong stellar IMF variations as a function of cluster mass are adopted, or 2) that lower mass clusters lose a significant fraction of all the enriched material processed through the polluting source (i.e., AGBs, FRMS, IBs).  The second option makes the already severe mass budget problem significantly worse.  Additionally, the observed inversion in the radial profiles of the primordial and enriched populations in M15 (and potentially other clusters) is not consistent with any of the self-enrichment scenarios put forward so far.

We conclude that none of the main sources of enrichment considered in the literature for the origin of the chemical anomalies in globular clusters are currently viable.  This, in turn, means that models that use these stellar sources to form enriched stars, either through secondary star-formation events (e.g., D'Ercole et al.~2008; 2010; Krause et al.~2013) or through the accretion of enriched material processed through such sources (e.g., Bastian et al.~2013b) do not appear to be viable.  

We hope that these results instigate future work on alternative scenarios that do not invoke nuclear burning as the origin of the chemical abundance anomalies (e.g., Hopkins~2014).  Additionally, work on whether stellar yields could be environmentally dependent in such a way as to reproduce the diversity of abundances (i.e. different He abundances for a given amount of Na and O) that are observed.  Finally, estimates for the He range present in a larger sample of clusters would further strengthen the constraints presented here.

\section*{Acknowledgments}

We thank Pavel Denisenkov for providing the yields for the VMSs, and for a helpful and critical reading of the paper.  We also thank Thibaut Decressin, Soeren Larsen, Henny Lamers and Carolyn Doherty for helpful discussions. The anonymous referee is thanked for helpful suggestions.  NB is partially funded by a Royal Society University Research Fellowship. 

\bsp
\label{lastpage}
\end{document}